\documentstyle[preprint,tighten,aps]{revtex}

\begin{document}

\draft

\title{Quark Delocalization, Color Screening Model and Nucleon-Baryon
Scattering}

\author{Guang-han Wu}
\address{Radiation Center, Institute of Nuclear Science and Technology,
Sichuan University, Chengdu, 610064, P. R.China}
\author{Jia-Lun Ping}
\address{ Department of Physics, Nanjing Normal University, Nanjing,
 210097, P. R.China}
\author{Li-jian Teng}
\address{Radiation Center, Institute of Nuclear Science and Technology,
Sichuan University, Chengdu, 610064, P. R.China}
\author{ Fan Wang}
\address {Center for Theoretical Physics and Department of Physics,
          Nanjing University, Nanjing, 210008, P. R. China.}
\author{ T. Goldman}
\address {Theoretical Division, T-5, Los Alamos National Laboratory,
 Los Alamos, NM87545, U.S.A }

\date{\today}
\maketitle

\vspace{-4.0in}
\flushright{LA-UR-98-5841}
\vspace{4.0in}

\begin{abstract}
We apply the quark delocalization and color screening model to
nucleon-baryon scattering. A semi-quantitative fit to N-N, N-$\Lambda$
and N-$\Sigma$ phase shifts and scattering cross sections is obtained
without invoking meson exchange. Quarks delocalize reasonably in all of
the different flavor channels to induce effective nucleon-baryon
interactions with both a repulsive core and with an intermediate range
attraction in the cases expected.
\end{abstract}

\flushleft

\newpage
\section{Introduction}
QCD has acquired significant experimental support as the correct
fundamental theory of the strong interaction. However in the low energy
region, its nonperturbative nature makes it hard to use directly for
the study of complicated systems such as hadron interactions and exotic
quark states. QCD inspired models are generally used in these cases.
Meson exchange models$^{\cite{deswart,Reuber}}$, on the other hand, are
quite successful in explaining the hadron interactions. But since the
quark and gluon degree of freedom are completely integrated out and
many phenomenological coupling constants and short-range phenomenology
are involved in this model, it is hard to use it to make predictions of
the properties of exotic quark states.  Hybrid meson-gluon exchange
quark cluster models have been developed by few
groups$^{\cite{Fujiwara,thomas,lomon}}$ and have achieved a
quantitative fit to the scattering data that accounts well for
nucleon-baryon interactions$^{\cite{Fujiwara}}$.  However, in
ref.\cite{Fujiwara} baryons are assumed to be so stiff that no internal
distortion would be induced, no matter how close the interacting
baryons are.  The Hamiltonian used is a direct extension of that used
in single hadrons.

It is possible that some physics has been precluded by these
assumptions a priori. For example, the QCD vacuum in between baryons
can be expected to vary as the quark matter density increases as two
colliding baryons approach each other and quark percolation between
hadrons may occur at short distances. On the other hand, the color
confinement interaction is screened at large distances due to
excitation of quark-antiquark ($q\bar{q}$) pairs, as has been shown by
unquenched lattice QCD calculation$^{\cite{Born}}$. Also, various kinds
of multigluon exchange interactions cannot be included in two body
confinement or the Fermi-Breit form for the interaction; nor can they
be studied in the pure valence quark model of single hadrons.  An
example of such a three gluon exchange interaction, which is impossible
for a $q\bar{q}$ meson and does not contribute within a colorless $q^3$
baryon but does contribute to hadron-hadron interactions and to
multiquark states, has been discussed in ref.{\cite{TGE}}.

Recently, the proposed $d'$ ($IJ^P=00^-$) dibaryon has been studied by
Faessler's group$^{\cite{Buchmann}}$, who concluded that both bag and
hybrid meson-gluon exchange models cannot obtain such a state with a
mass as low as 2.06GeV. If the $d'$ is proven to exist, that will argue
strongly against the completeness of these models. Moreover, some well
established facts are either impossible or hard to address within
either the meson exchange or the hybrid meson-gluon exchange model
approaches. For example, nuclear and molecular forces have been known
to be similar for more than half a century$^{\cite{Bohr}}$. If the
nuclear intermediate range attraction is due to meson exchange, then
the similarity would be accidental because the molecular force
certainly cannot be due to electron-positron pair exchange. The nucleus
as a collection of nucleons which has been proven to be a good
approximation through the nuclear structure studies, which leaves a
question as to why nucleus is not simply a collection of quarks.

A different approach has been developed by our group, which we call the
quark delocalization, color screening model (QDCSM)$^{ \cite{QDCSM}}$.
It is aimed at including more of the physics needed for the study of
exotic quark states. With the lessons$^{\cite{Isgur}}$ learned in
earlier studies of such states in mind, we first tested our model on
N-N scattering data$^{\cite{QDCSM}}$. We found that the QDCSM is able
to produce both the N-N short range repulsion and intermediate range
attraction simultaneously without invoking meson exchange. The
similarity between nuclear and molecular forces thus obtains a natural
explanation, i.e., the intermediate range attraction is due to
distortion of the internal structure of the constituent nucleons (and
atoms, respectively). Nuclear collections of 6$q$, 9$q$ and 12$q$
systems can be shown to be energetically favored when they have
structures organized into something close to the conventional form of
$d$, $^3H$, $^3He$ and $^4He$. On the other hand, a 6$q$ system with
$IJ^P=03^+$ quantum numbers is energetically favored to be in a six
quark state, $d^*$, instead of two $\Delta$s. This state is predicted
to be a narrow resonance with width of order MeV$^{\cite{Wong}}$.

Only very limited experimental data is available for nucleon-hyperon
(N-Y) scattering. Nevertheless it provides a further check of hadron
interaction models. Here we report the results of applying the QDCSM to
N-Y scattering. We should emphasize immediately that we are not so
ambitious as to expect to achieve a quantitative fit to all
nucleon-hyperon scattering with an almost parameter free model; more
modestly, we only seek to test further whether the QDCSM contains
basically the right physics.

\section{Quark delocalization and color screening model}
We present a two baryon system as an example to illustrate the QDCSM.
The generator coordinate method (GCM) is used to describe the 6$q$
system. The GCM basis wave function (WF) is assumed to be
\begin{eqnarray}
\Psi (1\cdots 6) & = & {\cal A}
[\psi_{B_1}(123)\psi_{B_2}(456)]_{ST},  \label{eq1} \\
\psi_{B_1}(123) & = & \chi_{c_1} (123) \eta_{S_1T_1}(123)\psi_L (1)
\psi_L (2)\psi_L (3), \label{eq1a} \\
\psi_{B_2}(456) & = & \chi_{c_1} (456) \eta_{S_2T_2}(456)\psi_R (4)
\psi_R (5)\psi_R (6), \label{eq1b}
\end{eqnarray}
where ${\cal A}$ is the normalized antisymmetrization operator and
$\chi_{c_i}, \eta_{S_iT_i} (i=1,2)$ are color, spin-isospin WFs.
$[~]_{ST}$ means that $S_1T_1, S_2T_2$ are coupled to channel spin $S$
and isospin $T$. $\chi_{c_i}$ is always the $3q$ color singlet state.
$\eta_{S_iT_i}$ is the $SU_6^{\sigma \tau}$ symmetric spin-isospin WF
for the N-N channel, but for $\Lambda$ and $\Sigma$, we use the $uds$
spin-flavor asymmetric hyperon WF,
\begin{eqnarray}
\Lambda \uparrow (456) & = & \sqrt{\frac{1}{2}} (u_4d_5-d_4u_5)s_6
 \sqrt{\frac{1}{2}} (\uparrow_4\downarrow_5-\downarrow_4\uparrow_5)\uparrow_6 
 \nonumber \\
\Sigma^0 \uparrow (456) & = & \sqrt{\frac{1}{2}} (u_4d_5+d_4u_5)s_6
 \sqrt{\frac{1}{6}} (2\uparrow_4\uparrow_5\downarrow_6
    -\uparrow_4\downarrow_5\uparrow_6-\downarrow_4\uparrow_5\uparrow_6), 
\label{eq2}
\end{eqnarray}
where the arrows refer to the quark and overall baryon spin, and
antisymmetrization will be applied to the five $u,d$ quarks only for
N-$\Lambda$ and N-$\Sigma$ channels. This choice explicitly
distinguishes $s$ quarks and allows for flavor symmetry breaking
effects to be calculated more easily, but the results are the same as
those which use an $SU_6^{\sigma f}$ symmetric baryon WF and totally
antisymmetric six quark states.

Our spatial WFs take the form 
\begin{eqnarray}
\psi_L({\vec r}) & = & \left( \phi_L({\vec r})+\epsilon \phi_R({\vec r})\right)
 /\sqrt{1+\epsilon^2 +2\epsilon \langle\phi_L|\phi_R\rangle}, \label{eq3a} \\
\psi_R({\vec r}) & = & \left( \phi_R({\vec r})+\epsilon \phi_L({\vec r})\right)
 /\sqrt{1+\epsilon^2 +2\epsilon \langle\phi_L|\phi_R\rangle}, \label{eq3b} \\
\phi_L({\vec r}) & = & \left(\frac{1}{\pi b^2} \right)^{3/4}
  e^{-\frac{({\vec r}-\vec{s}_1)^2}{2b^2}}, \label{eq3c} \\
\phi_R({\vec r}) & = & \left(\frac{1}{\pi b^2} \right)^{3/4}
  e^{-\frac{({\vec r}-\vec{s}_2)^2}{2b^2}}, \label{eq3d}
\end{eqnarray}
where $b$ is the size parameter of baryon.
$\vec{s}=\vec{s}_2-\vec{s}_1$, which is the separation of two reference
centers, plays the role of the generator coordinate in our model. The
only difference from the usual GCM basis WF is found in the single
quark orbital WF eq.(\ref{eq3a}, \ref{eq3b}):  A delocalization
parameter $\epsilon (s)$, which will be determined variationally by the
six quark dynamics for each separation $s = |\vec{s}|$, has been
introduced to describe the mutual percolation of quarks originally
confined in different baryons. This basis WF includes the six quark
bag-model-like WF ($\epsilon=1$ case, which is spherical only for
$\vec{s}=\vec{0}$) and the usual quark cluster model WF ($\epsilon=0$
case) as two extremes.  Note that intermediate configurations
corresponding to mutually distorted baryons are allowed within the
variational Hilbert space.

The Hamiltonian of the six quark system is taken as
\begin{eqnarray}
H(1\cdots 6) & = & \sum_{i=1}^6 (m_i+\frac{p_i^2}{2m_i}) +\sum_{i<j=1}^{6}
    \left( V_{ij}^c + V_{ij}^G \right) ,   \label{eq4} \\
V_{ij}^G & = & \alpha_s \frac{\vec{\lambda}_i \cdot \vec{\lambda}_j }{4}
 \left[ \frac{1}{r_{ij}}-\frac{\pi \delta (\vec{r})}{m_i m_j}
 \left( 1+\frac{2}{3} \vec{\sigma}_i \cdot \vec{\sigma}_j \right) \right],
  \label{eq5}
\end{eqnarray}
where $\vec{\lambda} (\vec{\sigma})$ is the $SU_3^c$ Gell-Mann
($SU_2^{\sigma}$ Pauli) operator, $\vec{r}=\vec{r}_i-\vec{r}_j$ and the
other symbols have their usual meaning.  Momentum dependent and tensor
interactions are neglected in this calculation. $V_{ij}^G$ is the
Fermi-Breit approximation to single gluon exchange and the color
screened confining interaction is defined by
\begin{equation}
V_{ij}^c= a_c \vec{\lambda}_i \cdot \vec{\lambda}_j \left\{
\begin{array}{ll}
 r & ~~~
\mbox{if }i,j\mbox{ occur in the same baryon orbit}, \\
\frac{1 - e^{-\kappa r} }{ \kappa } & ~~~
 \mbox{if }i,j\mbox{ occur in different baryon orbits}, \end{array} \right. \label{eq6}
\end{equation}
Explicitly, eq.(\ref{eq6}) means we use the non-screened,
color-confinement potential, namely linear confinement, to calculate
matrix elements $\langle LL |V|LL\rangle, \langle RR |V|RR \rangle
,\langle LL |V|RR\rangle$ and $\langle RR |V|LL\rangle$, and use the
color-screening confinement potential to calculate the other matrix
elements, such as $\langle LR |V|LR \rangle, \langle LL |V|LR\rangle$.
Here $\langle LL |V|LL\rangle = \langle \phi_L (i) \phi_L(j) |
V_{ij}|\phi_L(i)\phi_L(j) \rangle$, etc.

For $\langle LR |V|LR\rangle$, the interacting quarks are clearly
always in different baryons and so screened at longer distances.  For
$\langle LL |V|LR\rangle$, which form applies is not obvious and a
model choice must be made. One could consider other choices such as an
average of the two confinement forms. Another ambiguity occurs for
$s=0$, where we have only one baryon orbit. We define the matrix
elements at $s=0$ to be the limit of the values as $s$ tends to zero.
Both of these model choices are consistent. We recognize that making
these additional assumptions means that we no longer have simply a
potential model but rather that we are implicitly including an
approximation to many-body, low energy QCD interactions which cannot be
included in two body confinement and Fermi-Breit interactions. (A
similar inclusion occurs in models such as the "flip-flop"
model.~$^{\cite{yazaki}}$) In this sense, we are extending an effective
matrix element method from bound states to scattering states.

A good feature of this model Hamiltonian is that it reduces to the
usual non-screening, color-confinement model Hamiltonian for a single
hadron and for two hadrons in the asymptotic region but the spurious
color van der Waals force have been eliminated. The historical triumphs
of the constituent quark model in explaining hadron spectroscopy are
retained and the model parameters $m$ ($u,d$ quark mass), $\alpha_s$
(quark-gluon coupling constant), $a_c$ (strength of the confinement
potential) and $b$ (baryon size parameter) can be determined by the
nucleon mass, $N-\Delta$ mass splitting, and the stability condition
for nucleon size, $ \partial M_N(b)/\partial b =0$, along with the
usual choice, $m=M_N/3$. The $s$ quark mass ($m_s$) is most accurately
determined by the difference of $\Lambda$ and $\Sigma$ masses.

The color screening constant $\kappa$ is directly taken from lattice
QCD results$^{\cite{Born}}$ and this is the reason for our having
chosen a linear confinement and exponential color screening in this
calculation even though quadratic confinement may be more proper for a
nonrelativistic model$^{\cite{QDCSM}}$. The parameters fixed in this
way are:
\begin{eqnarray}
 & & m=313 \mbox{MeV}, m_s=521.7 \mbox{MeV}, b=0.625 \mbox{fm}, \nonumber \\
 & & \alpha_s=1.71, a_c=39.1 \mbox{MeV fm}^{-1}, \kappa=1.1111 \mbox{fm}^{-1} 
   \label{eq7}
\end{eqnarray}
Note that the $\Lambda$ and $\Sigma$ masses calculated are 1025 MeV and
1103 MeV, 90 MeV lower than experimental values.

\section{Calculation Method}
Due to delocalization, our GCM basis WF, eq.(\ref{eq1}), includes not
only the usual $q^3$-$q^3$ clustering, but also $q^6, q^5$-$q$ and
$q^4$-$q^2$ clustering, and therefore can not be factorized into
internal, relative and center of mass WF in the interaction region. The
usual cluster model method has to be extended$^{\cite{Wu}}$. Suppose
$\Psi$ is a solution of our model Hamiltonian eq.(\ref{eq4}),
\begin{equation}
(H-E) \Psi =0. \label{eq8}
\end{equation}

In general, both local and nonlocal interactions are included in $H$,
but the nonlocal interaction is nonzero only within a limited
interaction region $r < a$, where $a$ is roughly determined by the
scale at which the overlap of different orbitals, Eqs.(\ref{eq3c}) and
(\ref{eq3d}), becomes negligible, and $r$ refers here to the separation
between two three-quark clusters (see Eq.(\ref{eq16}) below).  If we
separate the whole space into interaction and asymptotic regions, then
in the interaction region, we can rewrite eq.(\ref{eq8})as,
\begin{equation}
(H+{\cal L}-E) \Psi = {\cal L} \Psi, \label{eq9}
\end{equation}
where ${\cal L}$ is the Bloch operator$^{\cite{Bloch}}$, which was
introduced by Bloch to make the Hamiltonian Hermitian in a finite
space. Taking the Hermitian conjugate of eq.(\ref{eq9}) and using
$H^{\dagger}+{\cal L}^{\dagger} = H + {\cal L}$, we obtain
\begin{equation}
\langle \Psi |H-E|\Psi_t \rangle |_0^a =
   \langle \Psi |{\cal L}^{\dagger}-{\cal L}|\Psi_t \rangle |_0^a.   \label{eq10}
\end{equation}
where the notation $|_0^a$ is intended to convey that the integration
in $r$ is restricted to $0 < r < a$ and $\Psi_t$ is a trial WF.  Then a
variational functional, $J(\Psi_t)|_0^a$, can be defined in the
interaction region as,
\begin{eqnarray}
J(\Psi_t)|_0^a & = & \langle \Psi_t | H-E |\Psi_t \rangle |_0^a
     - \langle \Psi | H-E |\Psi_t \rangle |_0^a \nonumber \\
& = & \langle \Psi_t | H-E |\Psi_t \rangle |_0^{\infty}
  -\langle \Psi_t | H-E |\Psi_t \rangle |_a^{\infty}
  -\langle \Psi | {\cal L}^{\dagger} - {\cal L} |\Psi_t \rangle |_0^a, \label{eq11}
\end{eqnarray}
and this functional does have a variational minimum with respect to
variation of the trial WF, $\Psi_t$. This can be seen from the fact
that the first line is quadratic in the difference between the trial
and exact WFs since, due to eq.(\ref{eq8}), $\langle \Psi_t | H-E |\Psi
\rangle |_0^a = \langle \Psi | H-E |\Psi \rangle |_0^a =0$.  The second
line simply involves rewriting the first term in terms of the full and
exterior ranges, and the second term has had the substitution made from
eq.(\ref{eq10}).

The GCM WF is written as
\begin{equation}
\Psi_t^{GCM} = \int f(\vec{s}) \Psi (\vec{s}) d\vec{s}, \label{eq12}
\end{equation}
where $\Psi(\vec{s})$ is the six quark WF of eq.(\ref{eq1}). Upon
substituting this trial WF in eq.(\ref{eq11}) and doing a partial wave
decomposition (Only central interactions are studied here; if
non-central interactions are to be included, this partial wave
decomposition would have to be extended correspondingly.), we obtain
\begin{equation}
J_l(\Psi_t)|_0^a = \int ds ds' f_l(s) f_l(s') \widetilde{K}_l^{GCM} (s,s')
   -{\cal L}_l(a), \label{eq13}
\end{equation}
where
\begin{eqnarray}
\widetilde{K}_l^{GCM} (s,s') & = & K_l^{GCM} (s,s')-K_l^{\prime GCM} (s,s'),
   \label{eq14}  \\
\frac{K_l^{GCM} (s,s')}{ss'} & = & \int Y_{lm}^* (\hat{s}) \langle \Psi (\vec{s})
 |H-E|\Psi (\vec{s}') \rangle Y_{lm}(\hat{s}') d\hat{s} d\hat{s}',
   \label{eq15}  \\
K_l^{\prime GCM} (s,s') & = & c_l \int_a^{\infty} dr \Gamma_l (r,s) \left[
  -\frac{\hbar^2}{2\mu}\left( \frac{d^2}{dr^2} -\frac{l(l+1)}{r^2} \right)+
    V^c(r)-E_r\right] \Gamma_l (r,s') \label{eq16}
\end{eqnarray}
where here, $\vec{r} = (\frac{\vec{r}_4+\vec{r}_5+\vec{r}_6}{3}-
  \frac{\vec{r}_1+\vec{r}_2+\vec{r}_3}{3}), ~c_l=1-(-)^{S+T+l}
  \delta_{AB}$, and $E_r$ is the energy of relative motion.

To obtain eq.(\ref{eq16}), we have assumed that, in the asymptotic
region: All of the exchange color interactions have died out; only the
long range Coulomb interaction, $V^c({r})$, for charged baryons may
remain; the delocalization has disappeared ($\epsilon=0$); and the six
quark system has clustered into two three quark baryons $A$ and $B$.

$\Gamma_l (r,s)$ is the $l$th partial wave of the relative motion WF,
$\Gamma (\vec{r},\vec{s})$, obtained from the WF of eq.(\ref{eq1}) when
it is factorized into internal, relative and center of mass parts of
two three quark clusters (see eq.(\ref{eq26}) below).
\begin{eqnarray}
\Gamma (\vec{r},\vec{s}) & = & \left( \frac{3}{2\pi b^2}\right)^{3/4}
 e^{-\frac{3}{4b^2}(\vec{r}-\vec{s})^2}, \label{eq17a} \\
\Gamma_l (r,s) & = & \left( \frac{3}{2\pi b^2}\right)^{3/4}
 e^{-\frac{3}{4b^2}(r^2+s^2)} 4\pi rs i^l j_l(-i\frac{3}{2b^2}rs). 
  \label{eq17b}
\end{eqnarray}

Finally
\begin{eqnarray}
{\cal L}_l(a) & = & (c_l \frac{\hbar^2}{2\mu}) \left[ g'_l(a)g_l^{t}(a)-g_l(a)g_l^{\prime t}(a)
  \right].  \label{eq18} \\
g_l^{t}(r) & = & \int f_l(s) \Gamma_l (r,s) ds, \label{eq19}
\end{eqnarray}
and $g_l(r)$ is the radial part of the relative motion WF of the exact
solution $\Psi$ of eq.(\ref{eq8}). The Bloch operator ${\cal L} =
\frac{\hbar^2}{2\mu}\frac{1}{a}\delta(r-a)\frac{d}{dr} r$ has been used
in deriving eq.(\ref{eq18}).

Next we follow Canto and Brink$^{\cite{Canto}}$, assuming as a boundary
condition that the trial and exact logarithmic derivatives are equal at
the boundary,
\begin{equation}
L^t_l = \frac{g_l^{\prime t}(a)}{g_l^{t}(a)} = L_l = \frac{g_l^{\prime}(a)}{g_l(a)}.
 \label{eq20}
\end{equation}
This implies that ${\cal L}_l(a)=0$. 

Making use of the stability property of the functional $J(\Psi^t)$, we
vary $J(\Psi^t)$ with respect to the trial WF $f_l(r)$ to obtain the
equation of motion of $f_l(r)$. The advantage of the Canto-Brink
variational method is that one need not solve this equation of motion,
but that instead one can obtain the logarithmic derivative $L_l^t$
directly via
\begin{eqnarray}
L_l^t & = & \frac{1}{\sum_{i,j} \Gamma_l(a,s_i) Q^{-1}_{ij} \Gamma_l(a,s_j)}
   \label{eq21} \\
Q_{ij} & = & \frac{2\mu}{\hbar^2}\frac{K_{ij}}{c_l} - \int_a^{\infty}
 dr \left[ \frac{d}{dr}\Gamma_l(r,s_i)\frac{d}{dr}\Gamma_l(r,s_j)+\Gamma_l(r,s_i)
 w(r) \Gamma_l(r,s_j) \right], \label{eq22} \\
K_{ij} & = & K_l^{GCM} (s_{i},s_{j}) \nonumber \\ 
w(r) & = & \frac{l(l+1)}{r^2} + \frac{2\mu}{\hbar^2} [ V^c(r)-E_r ]. \nonumber
\end{eqnarray}
Phase shifts can then be obtained through $L_l^t$,
\begin{equation}
\delta_l = \tan^{-1} \left[ \frac{kF'_l(ka)-L_l^t F_l(ka)}{L_l^tG_l(ka)-kG'_l(ka)}\right],
 \label{eq23}
\end{equation}
where $F$ and $G$ are Coulomb WFs. 

The difference between our derivation and that of Canto and Brink is
that we do not assume that the trial WF basis, $\Psi$ of
eq.(\ref{eq1}), can be factorized into the internal, relative and
center of mass parts in the interaction region. In fact, it is
impossible to do so for our delocalized WF.

The main task remaining is to calculate the GCM kernel $K^{GCM}
(\vec{s},\vec{s}')$. The color part is standard. The spin-isospin part
is also unaltered for the N-N channel, but different from that of
others$^{\cite{Oka}}$ for the N-$\Lambda$ and N-$\Sigma$ channels
because we use a spin-flavor asymmetric hyperon WF and the
antisymmetrization is restricted within the five $u,d$ quarks.
The orbital matrix elements are more involved due to
delocalization; in particular, the center of mass motion must be
properly eliminated. We use a momentum projection method to project out
the $\vec{P}_c=0$ part,
\begin{equation}
\frac{1}{\sqrt{V}} \int  d\vec{r}_c e^{-i\vec{P}_c \cdot \vec{r}_c}
  \Psi(\vec{r}_1,\ldots, \vec{r}_6). \label{eq24}
\end{equation}

As mentioned before, our GCM basis WF includes not only $l^3r^3$, but
also $l^6r^0$, $l^5r^1$, $l^4r^2$, $l^2r^4$, $l^1r^5$, $l^0r^6$
configurations. Here $l^{n_1}r^{n_2}$ means,
\begin{equation}
\prod_{i=1}^{n_1} \phi_L(\vec{r}_i-\vec{s}_1)
  \prod_{j=n_1+1}^{n_1+n_2} \phi_R(\vec{r}_j-\vec{s}_2)   \label{eq25}
\end{equation}
which can be factorized to the form
\begin{equation}
\Phi_{B_1}(\xi_1)\Phi_{B_2}(\xi_2) \exp \left\{ -\frac{1}{2b^2} \left[
\frac{n_1n_2}{n} (\vec{r}-\vec{s})^2 + n (\vec{r}_c-\vec{s}_c)^2 \right] \right\},
  \label{eq26}
\end{equation}
where
\begin{eqnarray}
\vec{r} & = & \vec{R}_2-\vec{R}_1, ~~~~\vec{s}=\vec{s}_2-\vec{s}_1, \nonumber \\
\vec{r}_c & = & \frac{n_1\vec{R}_1+n_2\vec{R}_2}{n}=\frac{\sum_{i=1}^n \vec{r}_i}{n},~~~
\vec{s}_c = \frac{n_1\vec{s}_1+n_2\vec{s}_2}{n}, \label{eq27} \\
\vec{R}_1 & = & \frac{\sum_{i=1}^{n_1} \vec{r}_i}{n_1},~~~
\vec{R}_2 = \frac{\sum_{j=n_1+1}^{n} \vec{r}_j}{n_2}, \nonumber  \\
\end{eqnarray}
and $\xi_1(\xi_2)$ are the internal coordinates of the $n_1(n_2)$ quark
cluster.  

Different configurations have different $\vec{s}_c$, but the same
$\vec{r}_c=(\sum_{i=1}^n \vec{r}_i)/n$. Let us introduce the parameter
center $\vec{t}=(\vec{s}_1+\vec{s}_2)/2$. Then
$\vec{s}_c=\vec{t}-(3-i)\vec{s}/6$, where $i=0,1,\ldots,6$ corresponds
to the $(n_1,n_2)=(6,0),(5,1),(4,2),(3,3),(2,4), (1,5),(0,6)$ particle
partitions. (This result can obviously be extended to any N-particle
system.) By means of $\vec{t}$ and $\vec{s}$, the exponential part of
eq.(\ref{eq26}) can be written as
\begin{equation}
\exp\left\{ -\frac{1}{2b^2} \left[ \frac{n_1n_2}{n}(\vec{r}-\vec{s})^2
  +n (\vec{r}_c-\vec{t})^2+n\frac{(3-i)^2}{36} s^2+\frac{n(3-i)}{3}
  (\vec{r}_c-\vec{t})\cdot \vec{s}\right] \right\}. \label{eq28}
\end{equation}

Due to the appearance of $\vec{r}_c$ only in the combination
$(\vec{r}_c-\vec{t})$ in eq.(\ref{eq28}), the momentum projection,
eq.(\ref{eq24}), can be written as
\begin{equation}
\frac{1}{\sqrt{V}} \int d\vec{r}_c e^{-i\vec{P}_c \cdot \vec{r}_c} 
  \Psi(\vec{r}_1,\ldots, \vec{r}_6) = \frac{1}{\sqrt{V}}
  \int d\vec{t} e^{-i\vec{P}_c \cdot \vec{t}}
  \Psi(\vec{r}_1,\ldots, \vec{r}_6) \stackrel{\vec{P}_c=0}{\longrightarrow}
\frac{1}{\sqrt{V}} \int d\vec{t} \Psi(\vec{r}_1,\ldots, \vec{r}_6)  \label{eq29}
\end{equation}
The spurious center of mass motion part of the GCM kernel can thus be
eliminated by a double momentum projection 
\begin{equation}
\frac{1}{V} \int d\vec{t} d\vec{t}' \langle \Psi(\vec{s}) | H-E |
   \Psi(\vec{s}') \rangle. \label{eq30}
\end{equation}
When we used this momentum projection method to calculate the matrix
elements of the kinetic energy and the Galilean noninvariant Darwin
term of the one gluon exchange Fermi-Breit interaction, we obtained the
same analytic formulas as those of Fujiwara$^{\cite{Fujiwara}}$. Our
variational method has been checked with Fujiwara's numerical
results$^{\cite{Fujiwara}}$ as well.

\section{Results}
Initially, we carried out a unified, parameter free (i.e., all
parameters are determined by single hadron properties and the color
screening constant is taken from lattice QCD,) model calculation of the
N-N, N-$\Lambda$ and N-$\Sigma$ interactions. This was done to check
whether the quarks delocalize reasonably in the different flavor
channels to give rise to qualitatively correct N-N, N-$\Lambda$ and
N-$\Sigma$ effective interactions. These effective interactions are
shown in figs.1--3 ($NN~~ST=01,10; N\Lambda~~ST=0\frac{1}{2},
1\frac{1}{2}, N\Sigma~~ST=0\frac{1}{2}, 1\frac{1}{2}, 0\frac{3}{2},
1\frac{3}{2}$) and indeed qualitatively reproduce the phenomenological
results. That is, the intermediate range attractions, usually assumed
to be due to meson exchange, are reproduced by the quark delocalization
in the QDCSM. In the N-N channels, this model even gives rise to
semi-quantitatively correct effective interactions; figs.4-5 show the
$^1S_0, ^3S_1$ and $^1D_2$ N-N phase shift fits.~\cite{arndt}

Figs.2-3 shows the results for $N$-$\Sigma$ and $N$-$\Lambda$
channels.  For $N$-$\Lambda$, the spin triplet state is somewhat more
attractive than the spin singlet; channel coupling adds a bit more
attraction to this state but leaves the spin singlet almost unchanged.
We note that $^{4}_{\Lambda}$H and $^{4}_{\Lambda}$He both have spin
zero ground states and spin one excited states. One might interpret
this as evidence that the spin-singlet $N$-$\Lambda$ interaction is
more attractive than the spin triplet. However, the situation is not so
transparent due to the complications of the interactions of the four
bodies involved and, in addition, $\Lambda$-$\Sigma^{0}$ mixing
effects. The spin one ground state of the deuteron is certainly an
indication that the spin triplet $N$-$N$ interaction is more attractive
than the spin triplet, and we might reasonably expect (by flavor
symmetry) that this should hold true for all octet-baryon combinations.
However, due to the strong tensor interaction from pion exchange, the
Nijmegen OBE model F~\cite{deswart} nonetheless includes a more
attractive spin singlet $N$-$N$ central interaction. Furthermore, there
is a paucity of direct data on scattering in the $Y$-$N$ channels, and
widely differing relative strengths for the central interaction are all
consistent with both the available data and the nuclear states referred
to above. The QDCSM, on the other hand, predicts that spin triplet
$N$-$N$ and $N$-$\Lambda$ interactions are stronger than spin singlet
ones.  Clearly, which central interaction is stronger in each case
merits additional study.

For $N$-$\Sigma$, we find the strongest attraction in the
$IJ=\frac{1}{2}1$ channel, while the $IJ=\frac{3}{2}0$, and
$\frac{1}{2}0$ channels both have a little weaker attraction (single
channel case), and the $IJ=\frac{3}{2}1$ channel is repulsive.  Channel
coupling has little effect on $\frac{1}{2}1$, and pushes $\frac{1}{2}0$
from attractive to repulsive. These show that the $N$-$\Sigma$
potential is more strongly spin and/or isospin dependent than the
$N$-$\Lambda$ potentials, which have a little weaker dependence on
spin.  These results are in qualtitative agreement with the
calculations of OBE models~\cite{deswart,Reuber} and hybrid quark model
calculations~\cite{Fujiwara}, except that our attraction for the
$N$-$\Sigma$($\frac{1}{2}1$) channel is too strong.

To check if one can obtain a semi-quantitative fit of N-$\Lambda$ and
N-$\Sigma$ scattering by fine tuning of the model parameters, two kinds
of adjustment of the color screening parameter, $\kappa$, have been
made:  The first keeps the color screening parameter $\kappa$ for the
$u,d$ quarks unchanged, i.e., $\kappa_u~=~\kappa_d~=~1.11$fm$^{-1}$,
but allows $\kappa_s$ for the $s$-quark to vary; the second one keeps
$\kappa_u~=~\kappa_d~=~\kappa_s~=~\kappa$ but allows the value of
$\kappa$ to vary for the N-$\Lambda$ and N-$\Sigma$ channels. Figs.6, 7
and 8 show phase shifts for the N$\Lambda~~ST=0\frac{1}{2},
1\frac{1}{2}$ and N$\Sigma~~ST=0\frac{1}{2}, 1\frac{1}{2}$ and
$0\frac{3}{2}, 1\frac{3}{2}$ channels with
$\kappa_{s}~=\frac{4}{9}~\kappa_u$.  These are similar, but not
identical to other hybrid quark model results$^{\cite{Fujiwara,15}}$.
Figs.9, 10, 11, 12 and 13 show integral and differential scattering
cross sections~\cite{hypdat,eis} for N$\Lambda, p\Sigma^{+}$ and
$p\Sigma^{-}$. A qualitative fit is obtained even though our model
phase shifts differ from others. This feature occurs in the meson
exchange model as well; the Nijmegen models D and F~\cite{deswart} also
have quite different phase shifts from each other.

Quantitatively, the N-$\Lambda$ total cross section is fit quite
well.~\cite{hypdat} We have shown that channel coupling does not change
the N-$\Lambda$ interaction very much; therefore, this good fit will be
maintained even after N-$\Lambda$ and N-$\Sigma$ channel couplings are
taken into account. The p-$\Sigma^{+}$ cross section found in the model
is larger than the experimental value~\cite{eis} and this channel does
not couple to any others. Hence, some fine-tuning may be needed.
Conversely, the p-$\Sigma^{-}$ cross section found in the model is
smaller than the experimental value.~\cite{eis} For that case, a
channel coupling calculation is needed to determine if the fit can be
improved by strong channel coupling effects.

\section{Conclusion}
The QDCSM has been used to calculate the effective N-N, N-$\Lambda$ and
N-$\Sigma$ interactions.  Linear confinement, with a color screening
constant taking from lattice QCD, has been used in this calculation.
All other model parameters are determined from the properties of
baryons. This means that we have a parameter-free model calculation for
nucleon-baryon (N-B) interations. We find that the quarks delocalize
reasonably in the different flavor channels (10 altogether) to induce
qualitatively correct, effective N-B interactions except that the
N-$\Lambda$ (1/2,1) and N-$\Sigma$ (1/2,1) channels have attractions
that are somewhat too strong. For the N-N channels, this model even
gives semi-quantitatively correct phase shifts in the $^{1}S_{0},
^{3}S_{1}$, and $^{1}D_{2}$  partial waves. After fine tuning the color
screening constant, we find it also gives semi-quantitatively correct
scattering cross sections for the N-$\Lambda$ and N-$\Sigma$ channels.

Several points need to be improved upon and to be checked further.

\begin{itemize}
\item{1.} So far, only single channel dynamical calculations have
been done. The effect of dynamical channel coupling must be checked, 
especially for the N-$\Sigma$ channel.

\item{2.} Only central interactions have been included in this
calculation. Non-central interactions need to be studied and higher
partial wave scattering should be checked correspondingly.

\item{3.} A better fit of the existing N-B scattering data better would
support the QDCSM, but achieving such an improvement is not be the
primary goal of this model calculation. In fact, nucleon-hyperon
scattering data is so sparse that different meson exchange models fit
the data perfectly well, and hybrid quark models can be made to fit the
data as well, if one is willing to fine tune as has been done for meson
exchange models.  Since the fundamental strong interaction theory is
certainly QCD, the N-B interactions should be an excellent area in
which to study non-perturbative QCD.  Meson-baryon and quark-gluon
descriptions both are able to describe the N-B interactions if we are
willing to include the whole hierachy of meson and baryon excited
states and the totality of quark-gluon interaction
diagrams$^{\cite{multi}}$. However, we may well ask which one is the
most economical approach for including major non-perturbative QCD
effects and for paving the way to new strong interaction physics such
as exotic quark-gluon states, strangelets and so on. That the QDCSM
obtains a qualitative, and in some cases, even a semi-quantitative, fit
to N-B scattering and few nucleon systems$^{\cite{QDCSM}}$ in its very
naive version, might be an indication that it includes a substantial
component of the true physics. The QCD basis of its model Hamiltonian
needs to be studied further.

\item{4.} Nucleon spin structure studies show that the pure valence
constituent quark model is only a first approximation. However,
quark-antiquark excitation Fock components are certainly present in the
ground state of the nucleon$^{\cite{qing}}$. This fact should be taken
into account in any quark model approach to hadron structure and hadron
interaction studies; the QDCSM also needs to be elaborated to include
it.

\end{itemize}

This research is supported in part by the Department of Energy under
contract W-7405-ENG-36 and in part by the NSF of China.

\pagebreak

\begin{center}
{\large {\bf FIGURE CAPTIONS}}
\end{center}

\noindent Fig.\ 1 ~~ Effective potential in MeV vs.~baryon separation in $fm$
for $N$-$N$ channels with\\
\hspace*{1.0in} confinement parameter value, $\kappa = 1.111 fm^{-1}$.

\noindent Fig.\ 2 ~~~ Effective potential in MeV vs.~baryon separation in $fm$
for $N$-$\Sigma$ channels with\\
\hspace*{1.0in} confinement parameter value, $\kappa = 1.111 fm^{-1}$.

\noindent Fig.\ 3 ~~~ Effective potential in MeV vs.~baryon separation in $fm$
for $N$-$\Lambda$ channels with\\
\hspace*{1.0in} confinement parameter value, $\kappa = 1.111 fm^{-1}$.

\noindent Fig.\ 4 ~~~ Phase shifts in degrees vs.~center of mass energy
in MeV for $^{1}S_{0}$ and $^{3}S_{1}$ $N$-$N$\\
\hspace*{1.0in} channels compared with data$^{\cite{arndt}}$.

\noindent Fig.\ 5 ~~~ Phase shifts in degrees vs.~center of mass energy
in MeV for $^{1}D_{2}$ $N$-$N$ channel\\
\hspace*{1.0in} compared with data$^{\cite{arndt}}$.

\noindent Fig.\ 6 ~~~ Scattering phase shifts in degrees vs.~center of
mass energy in MeV for $N$-$\Lambda$ channels\\
\hspace*{1.0in} with confinement parameter value, $\kappa_s = 
\frac{4}{9}\kappa $.

\noindent Fig.\ 7 ~~~ Scattering phase shifts in degrees vs.~center of
mass energy in MeV for $N$-$\Sigma$,\\ 
\hspace*{1.0in} $T=1/2$ channels with confinement parameter value, 
$\kappa_s = \frac{4}{9}\kappa $.

\noindent Fig.\ 8 ~~~ Scattering phase shifts in degrees vs.~center of
mass energy in MeV for $N$-$\Sigma$,\\ 
\hspace*{1.0in} $T=3/2$ channels with confinement parameter value, 
$\kappa_s = \frac{4}{9}\kappa $.

\noindent Fig.\ 9 ~~~ Cross section in $mb$ vs.~incident nucleon
laboratory momentum in MeV/$c$ for $N$-$\Lambda$\\
\hspace*{1.0in} scattering, compared with data$^{\cite{hypdat}}$.

\noindent Fig.\ 10 ~~ Cross section in $mb$ vs.~incident nucleon
laboratory momentum in MeV/$c$ for $p$-$\Sigma^{-}$\\
\hspace*{1.0in} scattering, compared with data$^{\cite{eis}}$.

\noindent Fig.\ 11 ~~ Cross section in $mb$ vs.~incident nucleon
laboratory momentum in MeV/$c$ for $p$-$\Sigma^{+}$\\
\hspace*{1.0in} scattering, compared with data$^{\cite{eis}}$.

\noindent Fig.\ 12 ~~ Differential cross section in $mb$ vs.~cosine of
center of mass scattering angle\\
\hspace*{1.0in} at incident $\Sigma^{-}$ laboratory momentum of 160 MeV/$c$ 
for $p$-$\Sigma^{-}$ scattering,\\
\hspace*{1.0in} compared with data$^{\cite{eis}}$.

\noindent Fig.\ 13 ~~ Differential cross section in $mb$ vs.~cosine of
center of mass scattering angle\\
\hspace*{1.0in} at incident $\Sigma^{+}$ laboratory momentum of 170 MeV/$c$ 
for $p$-$\Sigma^{+}$ scattering,\\
\hspace*{1.0in} compared with data$^{\cite{eis}}$.


\begin{references}
\bibitem{deswart} J.J.de Swart, P.M.M. Maessen and T.A. Rijken, in
	"Properties \& Interactions of Hyperons", eds. B.F. Gibson, 
	P.D. Barnes and K. Nakai, (World Scientific, Singapore, 1994), 
	p.37 and references therein.
\bibitem{Reuber} A.G. Reuber, {\it ibid.}, p.159 and references therein.
\bibitem{Fujiwara} Y. Fujiwara, C. Nakamoto and Y. Suzuki,
	Phys. Rev. Lett. {\bf 76}, 2242 (1996) and references therein.
\bibitem{thomas} K. Saito, K. Tsushima and A. W. Thomas, Nucl. Phys.
	{\bf A609}, 339 (1996).
\bibitem{lomon} P. LaFrance and E.L. Lomon, Phys. Rev. D {\bf 34}, 1341
	(1986).  
\bibitem{Born} K.D. Born, E. Lacrmann, N. Pirch, T.E. Walsh
	and P.M. Zerwas, Phys. Rev., D {\bf 40}, 1653 (1989).
\bibitem{TGE} W. Lucha, F.F. Sch\"{o}berl and D. Gromes, Phys. Rep.
	{\bf 200}, 127 (1991).  \bibitem{Buchmann} A.J. Buchmann, G.
	Wagner and A. Faessler, Phys. Rev. C {\bf 57}, 3340 (1998).  
\bibitem{Bohr} A. Bohr and B.R. Mottelson, Nuclear Structure, Vol.I,
	(W.A. Benjamin, Inc., New York, 1965), p.269.  
\bibitem{QDCSM} T. Goldman, in "Nuclear Chromodynamics", eds. S. Brodsky
	and E. Moniz, (World Scientific, Singapore, 1986), p.363. \\ 
	T. Goldman, K. Maltman, G.J. Stephenson, Jr. and K.E. Schmidt,
	Nucl. Phys. {\bf A481}, 621 (1988). \\ 
	F. Wang, G.H. Wu L.J. Teng and T. Goldman, Phys. Rev. Lett. 
	{\bf 69}, 2901 (1992). \\ 
	K. Maltman, G.J. Stephenson, Jr. and T. Goldman, Phys.
	Lett.  {\bf B324}, 1 (1994). \\ 
	F. Wang, J.L. Ping, G.H. Wu, L.J. Teng and T. Goldman, Phys. 
	Rev. C {\bf 51}, 3411 (1995). \\ 
	G.H. Wu, L.J. Teng, J.L. Ping, F. Wang and T. Goldman, Phys.
	Rev.  C {\bf 53}, 1161 (1996). \\ 
	T. Goldman, K. Maltman, G.J. Stephenson, Jr., J.L. Ping and 
	F.  Wang, Mod. Phys. Lett. {\bf A13}, 59 (1997).  
	\bibitem{Isgur} N. Isgur, in "Topical Conference on Nuclear 
	Chromodynamics", eds. J. Qiu and D. Sivers, (World Scientific, 
	Singapore, 1988), p.201.  
\bibitem{Wong} C.W. Wong, Phys. Rev. C {\bf 57}, 1962 (1998).  
\bibitem{yazaki} F. Lenz, J.T. Londergan, E.J. Moniz, R. Rosenfelder,
	M. Stingl and K. Yazaki, Ann. Phys. {\bf 170}, 65 (1986).
\bibitem{Wu} G.H. Wu and D.L. Yang, Sci. Sinica, Supp. A, 1112 (1982).  \bibitem{Bloch} C. Bloch, Nucl. Phys. {\bf 4}, 503 (1957).  
\bibitem{Canto} L.F. Canto and D.M. Brink, Nucl. Phys. {\bf A279}, 85 (1977).  \bibitem{Oka} M. Oka, K. Shimizu, and K. Yazaki, Nucl. Phys. {\bf A464}, 700
	(1987).  
\bibitem{15} U. Straub, {\it et al.}, Nucl. Phys. {\bf A483}, 686 (1988); 
	{\bf A508}, 385c (1990).  
\bibitem{multi} H. Hofestadt, S. Merk and H.R. Petry, Z. Phys. {\bf
	A326}, 391 (1987). \\
	F. Wang, Prog. Phys. {\bf 9}, 297 (1989). \\
	F. Wang and C.W. Wong, in "Quark-Gluon Structure of Hadrons and 
	Nuclei", eds. L.S. Kisslinger and X.J. Qiu (International Academic 
	Press, London, 1991), p.100. \\
	R.L. Jaffe, Nucl. Phys. {\bf A522}, 365c (1991).
\bibitem{qing} D. Qing, X.S. Chen and F. Wang, Phys. Rev. C {\bf 57},
	R31 (1998); Phys. Rev. D {\bf 58} (in press).
\bibitem{arndt} C. Oh, R.A. Arndt, I.I. Strakovsky and R.L. Workman, 
	Phys. Rev. C {\bf 56}, 635 (1997); NUCL-TH/9702006; R.A. Arndt, J.S. 
	Hyslop, III, and L. D. Roper, Phys. Rev. D {\bf 35}, 128 (1987); 
	R.A. Arndt {\it et al.}, Phys. Rev. D {\bf 28}, 97 (1983).
\bibitem{hypdat} G. Alexander, U. Karshon, A. Shapira, G. Yekutieli, 
	R. Engelmann, H. Filthuth and W. Lughofer, Phys. Rev. {\bf 173}, 
	1452 (1968); B. Sechi-Zorn, B. Kehoe, I. Twitty and 
	R.A. Burnstein, Phys. Rev. {\bf 175}, 1735 (1968); J.A. Kadyk, 
	G. Alexander, J.H. Chan, P. Gaposchkin and G.H. Trilling, Nucl. 
	Phys. {\bf B27}, 13 (1971).
\bibitem{eis} F. Eisele, H. Filthuth, W. F\"{o}hlish, V. Hepp and G. Zech, 
	Phys. Lett. {\bf B37}, 204 (1971). 
\end{references}
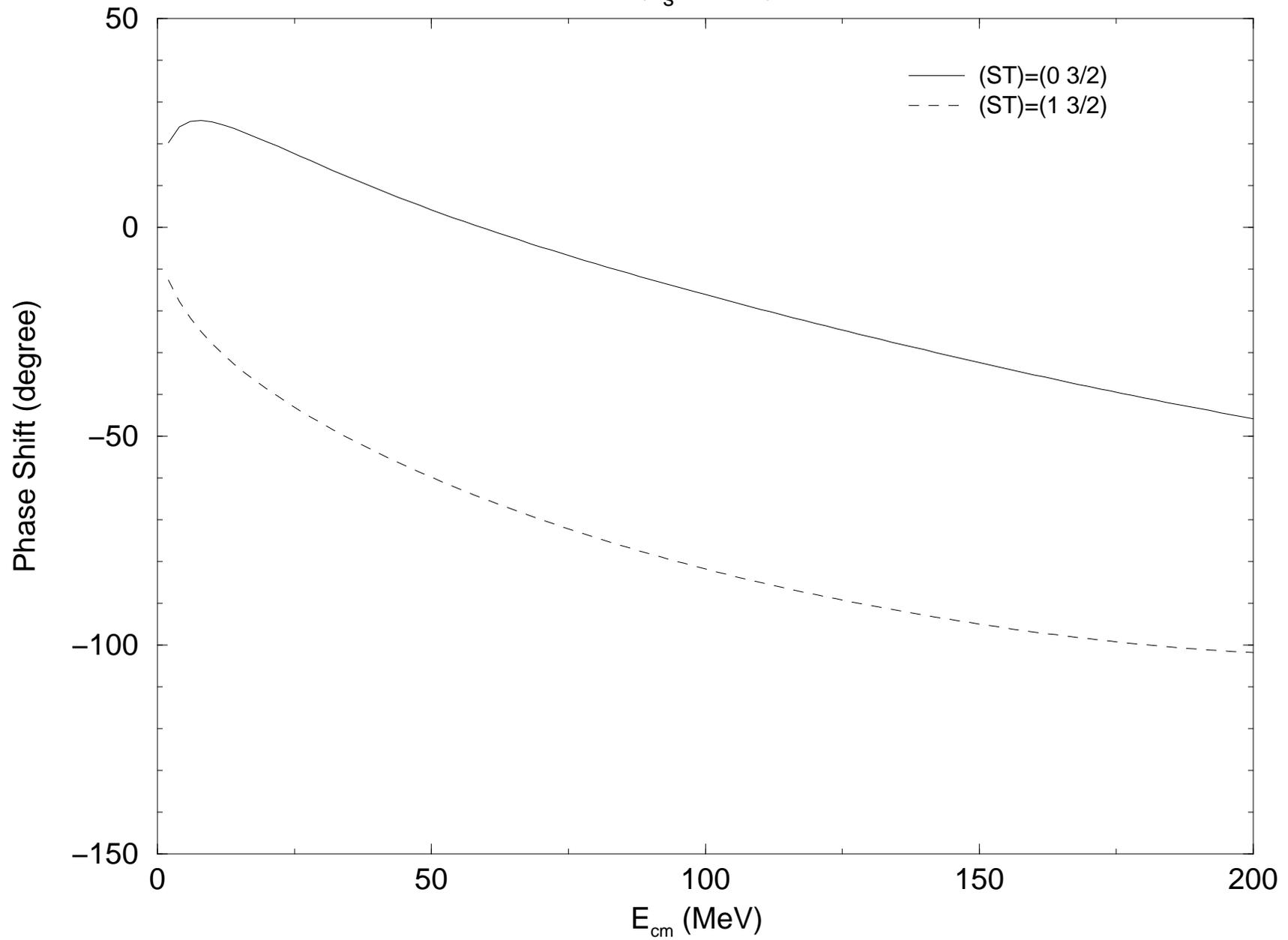
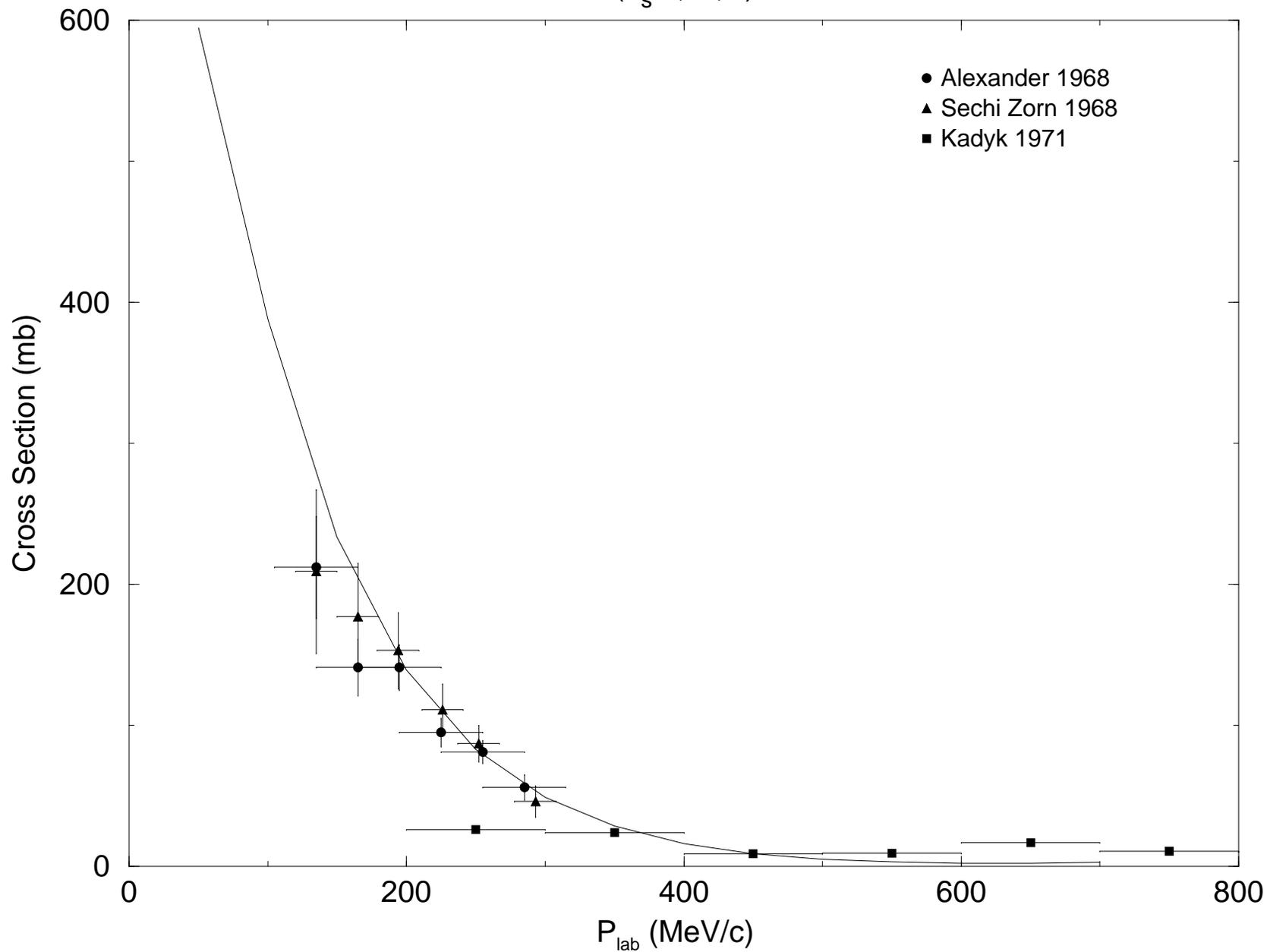
\end{document}